\documentclass[aps,preprint,amsmath,amssymb]{revtex4}
\usepackage{graphicx, color}

\def\be{\begin{eqnarray}}
\def\ed{\end{eqnarray}}
\def\non{\nonumber}
\def\la{\langle}
\def\ra{\rangle}
\def\UP{\cal U}
\def\shro{Schr$\rm\ddot{o}$dinger~}

\begin{document}

\preprint{IPMU09-0107}

\title{ \bf
Sommerfeld Enhancement from Unparticle Exchange \\
for Dark Matter Annihilation }

\vspace{5cm}

\author{\bf
Chuan-Hung Chen$^{1,2}$\footnote{E-mail: physchen@mail.ncku.edu.tw
},~~ C. S. Kim$^{3,4}$\footnote{E-mail: cskim@yonsei.ac.kr} }
\address{
$^{1}$ Department of Physics, National Cheng-Kung University, Tainan 701, Taiwan \\
$^{2}$National Center for Theoretical Sciences, Hsinchu 300, Taiwan
\\
$^{3}$ Department of Physics $\&$ IPAP, Yonsei University, Seoul 120-479, Korea \\
$^{4}$ IPMU, University of Tokyo, Kashiwa, Chiba 277-8568, Japan
}

\begin{abstract}
\noindent We investigate the implication of unparticle exchange for
the possible Sommerfeld enhancement in dark matter annihilation
process. Assuming the unparticle exchange during WIMP collision, we
solve the  Schr$\rm\ddot{o}$dinger equation for the effective
potential, and find that
the Sommerfeld enhancement factor is dictated by the scale dimension of unparticle as
$1/v^{3-2d_{\UP}}$. Numerically the Sommerfeld enhancement
could be ${\cal O}(10-10^{3})$.

\end{abstract}

\maketitle

\newpage
\noindent[{\it Introduction}]~~~~
It has been known that our universe is made of not only the stuff of
the standard model (SM) with the occupancy of mere $4\%$, but also dark
matter and dark energy with the abundance of $22\%$ and $74\%$,
respectively~\cite{PDG08}. Therefore, it must be one of the most important issue to
understand what the dark matter is and how to explore it by various
observations. Hopefully, through high energy colliders such as the
Large Hadron Collider (LHC) at CERN, we may directly observe dark
matter soon. In the mean time we may also have the
chance to probe dark matter indirectly by the study of the high energy cosmic-ray.

Recently, the collaborations of PAMELA~\cite{PAMELA},
ATIC~\cite{ATIC}, FERMI-LAT~\cite{Fermi},  HESS~\cite{HESSnew} and
$etc.$ have published quite astonished events in cosmic-ray
measurements, in which PAMELA observes the excess in the positron
flux ratio over 10 GeV till PAMELA's observational limit of about
100 GeV, whereas others measure consistent anomalies in the
electron+positron flux in the $300-1000$ GeV range.
Inspired by the new founds at the satellite, balloon and ground-based experiments,  the
excess could be readily ascribed to dark matter annihilation,  even
though there still are the possibilities of existing new young
pulsars \cite{pulsar}. Although dark matter $decays$ could be the
origin of such anomaly, however, for making the lifetime as long as
${\cal O}(10^{25})$ seconds, the extreme fine-tuning on the coupling of
interaction \cite{CGZ_PLB,Chen} cannot be avoided. For escaping the
fine-tuning problem, hereafter, we will focus on the mechanism of
dark matter $annihilation$ only. In addition,  the candidate of dark
matter in our following analysis is regarded as weakly interacting
massive particle (WIMP).

Although the WIMP annihilation could be the source for the excess of
cosmic-ray, however, due to low reacting rate in the annihilating
process, an enhanced boost factor of a few orders of magnitude,
$e.g.$ Sommerfeld enhancement
\cite{So0,So1,So2,So3,So4,So5,So6,So7}, has to appear during the
annihilation of WIMP. Therefore, a new force carrier in the dark
matter annihilation to dictate the enhancement is required. As an
example, an interesting mechanism for the Sommerfeld enhancement is
arisen from the light boson exchange between dark matter
\cite{So4,CGW}, where the resulted interaction is Yukawa potential
and the force carrier has the significant influence in the range of
Compton wavelength, denoted by $\alpha M_{\chi}$ with $\alpha$ being
the fine structure constant of the interaction and of order
$10^{-2}$.

\begin{figure}[bpth]
\includegraphics*[width=4 in]{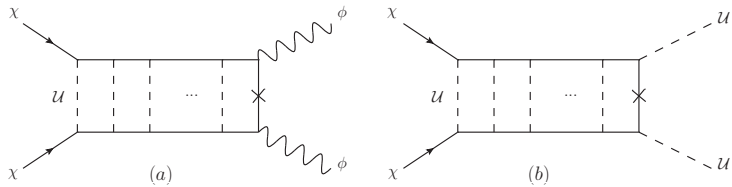}
\caption{Dark matter annihilation with unparticle-mediated
Sommerfeld enhancement. }
 \label{fig:unparticle}
\end{figure}

In this Letter, we study another kind of new force that may be alive in an
invisible sector and dictated by scale invariant. As known that an
exact scale invariant stuff cannot have a definite mass unless it is
zero, therefore for distinguishing from the conventional particles,
Georgi named the stuff as unparticle \cite{Georgi1, Georgi2}.
Interestingly, it is found that the unparticle with the scaling
dimension $d_{\UP}$ behaves like a non-integral number $d_{\UP}$ of
invisible particles \cite{Georgi1}. Further implications of the unparticle to
colliders and low energy physics could be referred to
Refs.~\cite{un1,un2,un3}.
In order to concentrate on the Sommerfeld effect,
here we don't study the general effective interactions with unparticle,
$e.g.$  in our analysis, we have suppressed the interactions between unparticle and Higgs \cite{Fox:2007sy}.
Although WIMP and unparticle both weakly
couple to the SM particles, however, there is no any reason to limit
that the interactions between them should be weak.
Therefore, 
when WIMPs collide each other with small speed, we speculate that
the Sommerfeld enhancement could be arisen from the unparticle
exchange during the collision, sketched in
Fig.~\ref{fig:unparticle}, where the $\chi$ and ${\cal U}$ denotes
the WIMP and unparticle, respectively. And $\phi$ represents the
(generic light) particle that might weakly decay to SM particles and
the constraints on the couplings will be controlled by current
observed fluxes of cosmic rays such as electrons, positrons,
antiproton and $etc.$ Here, its appearance is responsible for the
possible connection between dark and visible sectors, but not for
the Sommerfeld enhancement. Hence, we don't further discuss the
detailed couplings to the SM stuff and the related issue for this
decay. Our motivation is to understand that if there exists
unparticle in the invisible sector, what are its unique character on
the Sommerfeld factor and the differences from Coulomb and Yukawa
interactions?

The paper is organized as follows: First, we derive
the static unparticle potential, solve the associated radial \shro
equation with suitable boundary condition, and find the resultant
formula for $s$-wave Sommerfeld enhancement. Then,
we do the numerical analysis on the resulted Sommerfeld factor and
present it as a function of involved parameters in two-dimensional
contour plots. Finally we give the conclusion.
\\

\noindent[{\it Unparticle potential and its Sommerfeld
factor}]~~~~
Although the nonperturbative Sommerfeld effect can be calculated by
the combined contributions of a set of ladder diagrams shown as in
Fig.~\ref{fig:unparticle}, however in the non-relativistic limit,
the effect could be equivalent to solving the \shro equation with an
effective potential, which is arisen from the single particle
exchange. In the considered mechanism, here the exchanged particle
is the unparticle.
Following the scheme proposed in Ref.~\cite{Georgi1}, the
interaction of WIMP to unparticle can be written as
 \be
\frac{\lambda}{\Lambda_{\UP}^{d_{\UP} -1} }\bar \chi^{(c)}  \chi
{\cal O}_{\UP} \label{eq:unint}
 \ed
where we have assumed that the WIMP is a Dirac (or Majorana)
fermion, $\lambda$ is dimensionless parameter and $\Lambda_{\UP}$
denotes the living scale of unparticle. For displaying the character
of scale invariant stuff, we concentrate only on the scalar
unparticle.  Please note that when unparticle  is realized in the
framework of conformal field theories, the propagators for vector
and tensor unparticles should be modified appropriately to preserve
the conformal symmetry \cite{Grinstein:2008qk}. To obtain the
unparticle potential in non-relativistic limit, we use the
propagator of the scalar unparticle operator given by
\cite{Georgi1,Georgi2}
 \be
 \int e^{iq x}
 \la 0|
T{\cal O}_{\UP}(x) {\cal O}_{\UP}(0)|0\ra &=& i
\frac{A_{d_{\UP}}}{2\sin d_{\UP}\pi}
\frac{1}{\left(- q^{2}\right)^{2-d_{\UP}}}\,,\label{eq:unpro} \\
{\rm where}~~~~A_{d_{\UP}}&=& \frac{16 \pi^{5/2}}{(2\pi)^{2d_{\UP}}}
\frac{\Gamma(d_{\UP}+1/2)}{\Gamma(d_{\UP}-1) \Gamma(2
d_{\UP})}\,.\non
 \ed
Combining Eqs. (\ref{eq:unint}) and (\ref{eq:unpro}),
the four-fermion effective interacting term in momentum
space could be expressed by
 \be
\bar\chi^{(c)} \chi \left[\left( \frac{\lambda}{\Lambda^{d_{\UP}
-1}_{\UP}}\right)^2  \frac{A_{d_{\UP}}}{2\sin d_{\UP} \pi}
\frac{1}{(-q^2)^{2-d_{\UP}}}\right] \bar\chi^{(c)}\chi\,.
\label{eq:un_int}
 \ed
By Fourier transformation and with $q^0=0$, the static unparticle
potential in WIMP interaction resulted by Eq.~(\ref{eq:un_int}) is
found by
 \be
V(r)&=&  - \frac{\alpha}{r^{t}} \label{eq:unV}
 \ed
with $t = 2d_{\UP}-1$,
 \be
\alpha&=&\frac{\xi_{\Gamma}}{2\pi^{2d_{\UP}}}  \left(
\frac{\lambda}{\Lambda_{\UP}^{d_{\UP}-1}}\right)^2\,, \non \\
\xi_{\Gamma} &=& \frac{\Gamma(d_{\UP}+1/2)
\Gamma(d_{\UP}-1/2)}{\Gamma(2d_{\UP})}\,. \label{eq:coeff}
 \ed
Interestingly, we see that the power of unparticle potential is
associated with the scaling dimension $d_{\UP}$ and not an integer.

Since the unparticle potential is independent of the polar and
azimuth angles in spherical coordinate system,
the relevant piece for the Sommerfeld factor is the radial \shro
equation, read by
 \be
\left[\frac{d^2}{dr^2} + \left( -\frac{\ell(\ell+1)}{r^2}
+\frac{2\mu\alpha}{ r^t} + k^2 \right)\right]u_{k\ell}(r)=0\,,
\label{eq:rad}
 \ed
where we have used the nature unit with $\hbar=c=1$, $E=k^2/2\mu$
with $\mu$ being the reduced mass of system and
$R_{k\ell}(r)=u_{k\ell}(r)/r$. Once we solve the differential
equation with the proper boundary condition, the Sommerfeld effect
associated with each angular momentum $\ell$ is obtained by
\cite{So6}
 \be
S_{\ell}&=& \left| \frac{(2\ell+1)!!}{|k|^{\ell}\ell!}
\frac{\partial^{\ell} R_{k\ell}(r) }{\partial r^{\ell}}
\Big|_{r=0}\right|^2 \,. \label{eq:SF}
 \ed
Due to the power $t$ of unparticle potential in $r$ being
not an integer, there is no hope to find a general close form for the
solution. Nevertheless, since the required Sommerfeld factor is
estimated at $r=0$, our strategy for finding the solution is to look
for a good approximation to extract the behavior of radial wave
function at $r\sim 0$.

Before discussing the solution to the differential equation, first
we analyze the limit on $t=2d_{\UP}-1$.
To avoid the crossing of the branching cut at $1/\sin d_{\UP}\pi$ in
Eq.~(\ref{eq:unpro}),
we require $1 < d_{\UP} < 2\ (1<t<3)$.
In order to further understand whether the upper limit of $t$ can be
bounded by the boundary condition when solving differential
equation, we examine the case with $t=2$, in which the potential
could be exactly solved at $r\to 0$. Hence, by taking $t=2$ and
$u_{k\ell}\sim r^{\sigma}$ and considering $r\to 0$, from
Eq.~(\ref{eq:rad}) we get
 \be
\sigma=\frac{1}{2}+\sqrt{(\ell+\frac{1}{2})^2-2\mu\alpha}\,.\label{eq:lambda}
  \ed
Thus,
the corresponding solution for $u_{k\ell}(r)$ can be expressed by
 $u_{k\ell}(r)=r^{\sigma} e^{ikr} f(r)$,
where the function of $f(r)$ is controlled by
 \be
r\frac{d^2f}{dr^2}+(2\sigma +2ikr) \frac{df}{dr} + 2ik\sigma
f(r)=0\,. \label{eq:fr}
 \ed
Compare to the confluent hypergeometric differential equation,
 $x y''+(c-x) y'-a y=0$,
in which the solution is confluent hypergeometric function of the
first kind $ y(x)=  {}_1F_1(a,c;x)\,,$ we see that the solution to
Eq.~(\ref{eq:fr}) can be found by taking $a=\sigma$, $c=2\sigma$
and $x=-2ikr$.
For $\ell=0$, from Eq.~(\ref{eq:lambda}) it is easy to find
$\sigma< 1$.
Since ${}_1F_1(\sigma,2\sigma,-2ikr)\to 1$ when $r\to 0$, as a
result, the wave function $R_{k\ell}(r)$ is singular at origin. In
other words, $t=2$ leads to an ill-defined solution and should be
taken as the new upper limit. Therefore, we find the new bound on
the scale dimension of unparticle as $1<d_{\UP} <3/2\, (1<t<2)$.

After obtaining the allowed power of unparticle potential, next we
are looking for a good approximation to get the proper solution for
near origin. As used before, if we regard $u_{k\ell}(r)=r^{\ell+1}
e^{ikr} f(r)$ as a general solution to the radial \shro equation,
according to Eq.~(\ref{eq:rad}), the function $f(r)$ should satisfy
the differential equation
 \be
r \frac{d^2f}{dr^2} &+& (2(\ell+1)+2ikr) \frac{df}{dr}
 +
\left( 2ik(\ell +1) + \frac{2\mu\alpha}{ r^{t-1}}\right) f=0\,.
\label{eq:t12}
 \ed
Although this equation cannot be solved for arbitrary value of $t$,
we find that if we concentrate on small $r$, the terms associated
with $2ikr$ and $2ik(\ell+1)$ can be safely neglected and the
solution to the simplified differential equation can be found by
 \be
f_t(r)&=&
%
d^{-(2\ell+1)/2(2-t)} (2-t)^{(2\ell +1)/(2-t)}
\frac{1}{r^{(2\ell +1)/2}} \non\\
&\times& \Gamma\left( \frac{2\ell-t+3}{2-t}\right)
J\left(\frac{2\ell+1}{2-t}, \frac{2\sqrt{d}r^{(2-t)/2}}{2-t}
\right)\,, \non
 \ed
where $J_\nu(x)=J(\nu, x)$ is the Bessel function and the asymptotic
form for $0<x\ll \sqrt{\nu+1}$ is
 $J_\nu\to \left( x/2\right)^{\nu}/ \Gamma(\nu+1)$.
Hence, the solution to radial \shro equation in the region of small
$r$ can be expressed by
 \be
R_{k\ell}(r)=\frac{u_{k\ell}(r)}{r}=r^{\ell} e^{ikr} f_t(r)\,,
\label{eq:Rt12}
 \ed
where a suitable normalization for the wave function has to be
further dealt with.

Unlike conventional approach that the normalization of wave function
for scattering process is chosen at $r\to \infty$, in order to get a
suitable normalization for the wave function at near origin, we have
to consider an alternative method which could provide correct value
of wave function at $r=0$.
We find that the purpose can be achieved by using integral form
instead of directly solving the Eq.~(\ref{eq:rad}). In other words,
the solution for $u_{k\ell}(r)$ could be expressed by
 \be
u_{k\ell}(r)&\propto& j_{\ell}(kr) + \int^{\infty}_{0} dr' G(r,r')
U(r') u_{k\ell}(r') \label{eq:int_u}
 \ed
with $U(r)=2\mu V(r)$ and
 \be
 G(r,r')&=& \left\{
              \begin{array}{c}
                 j_{\ell}(kr) n_{\ell}(kr')/k \ \ \ {\textrm{for $r < r'$} }\,, \\
                 n_{\ell}(kr) j_{\ell}(kr')/k \ \ \ {\textrm{for $r >r'$} }\,, \\
              \end{array}
            \right. \label{eq:green}
 \ed
where
$G(r,r')$ is the Green's function of the differential equation.
After compared to the partial wave expansion given by
$\exp(ikz)=  \sum^{\infty}_{\ell=0}i^{\ell} (2\ell+1) j_{\ell}(kr)
P_{\ell}(\cos\theta)/kr$,
we find that the solution for $r\to 0$ is
 \be
u_{k\ell}= \frac{j_{\ell}(kr)}{k} +
\frac{j_{\ell}(kr)}{k^2}\int^{\infty}_{0} dr' n_{\ell}(kr')
j_{\ell}(kr')U(r')\,. \label{eq:ukl}
 \ed

Now we  match both solutions at
$r\approx 0$ by solving differential equation and the integration
with Green function. It has been known that the Sommerfeld factor of
Coulomb potential ($i.e.$ $t=1$) for $s$-wave can be obtained exactly as
 \be
 S_0 &=& \frac{2\pi\alpha/v}{1-\exp(-2\alpha\pi/v)}\,.
 \ed
For the case of $v\ll 1$, we have $S_0 \approx 2\pi \alpha/v$. Based
on the known result, we find that in order to get correct
approximation for Coulomb potential, matching condition for
Eqs.~(\ref{eq:Rt12}) and (\ref{eq:ukl}) has to be adopted as
 \be
N^2_{\ell} r^{\ell+1} f(r)= \frac{
4}{k^2}j_{\ell}(kr)\int^{\infty}_{0} dr' n_{\ell}(kr')
j_{\ell}(kr')U(r')\,.\non
 \ed
Using the asymptotic results
 $f(r\to 0)\longrightarrow 1$
and
 $j_{\ell}(kr\to 0)\to \frac{2^{\ell} \ell!}{(2\ell+1)!}
(kr)^{\ell+1}$,
the normalization constant is found by
 \be
N^2_{\ell} &=&  8\mu\alpha \frac{2^{\ell} \ell!}{(2\ell+1)!}
k^{\ell+t-2} X_{\ell}
  \ed
with $ X_{\ell}=\int^{\infty}_{0} dz j_{\ell}(z) n_{\ell}(z) / z^t$.
According to Eq.~(\ref{eq:SF}), the Sommerfeld factor for $s$-wave
is obtained by
$S_{0}= 8\alpha | X_0|\mu^{t-1}/ v^{2-t}$,
where we have used $k = \mu v$. One can see that for $t=1$,
$X_0=-\pi/4$ and $S_{0} =2\pi \alpha/v$.
 Hence, the $s$-wave Sommerfeld
factor induced by unparticle exchange is given by
 \be
S_{0}=\left(\frac{2\lambda}{\pi^{d_{\UP}}} \right)^2 \xi_{\Gamma}
\left( \frac{\mu}{\Lambda_{\UP}}\right)^{2(d_{\UP}-1)}
\frac{|X_0|}{v^{3-2d_{\UP}}}\,. \label{eq:S0}
 \ed
\\

\noindent[{\it Numerical analysis and discussions}]~~~~
We now analyze the formula for $S_0$ numerically. From
Eq.~(\ref{eq:S0}), we see that the involved free parameters are
$\lambda$, $d_{\UP}$, $\Lambda_{\UP}$ and $v$. Unlike the case of
Coulomb or Yukawa potential, the velocity-dependent factor appears
by  $1/v^{3-2 d_{\UP}}$ with $1< d_{\UP} < 3/2$. The Sommerfeld
factor is increasing when $d_{\UP}$ approaches to unity.
Consequently, if we focus on the maximum of $S_0$, then we find that
$S_0$ is insensitive to the values of $\mu$ and $\Lambda_{\UP}$,
where they show up by $(\mu/\Lambda_{\UP})^{2(d_{\UP}-1)}$.
 Note that to explain the excess of cosmic rays from both PAMELA and ATIC/Fermi-LAT, it has been known  that the mass of dark matter is of order of TeV.
 On the other hand, in order to produce the unusual stuff at LHC, the interesting scale
 to form unparticle should be also at TeV scale. Therefore,
without
lose of generality
we set $\mu=\Lambda_{\UP}$ in our numerical estimates.

In order to see the influence of remaining parameters, we will fix
one parameter in turn when we make two-dimensional contours as a
function of remaining two parameters. First, we analyze the contour
as a function of $\lambda$ and $d_{\UP}$ when the speed of WIMP is
fixed.
Although dark stuff weakly couples to visible particle, however, the
interaction between invisible particles may not be small.
\begin{figure}[t]
\includegraphics*[width=3 in]{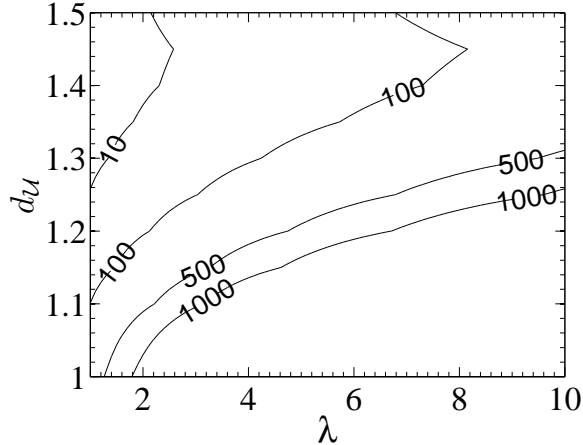}
\caption{Contour for Sommerfeld factor induced by
unparticle-mediated as a function of $\lambda$ and $d_{\UP}$ with
$v=10^{-3}$. The numbers in the plot denote the values of $S_0$. }
 \label{fig:sold}
\end{figure}
Accordingly, we set the allowed range for $\lambda$ be within one
order of magnitude, $i.e.$ $1<\lambda <10$. As a result, the contour
for Sommerfeld factor $S_0$ induced by unparticle contributions as a
function of coupling $\lambda$ and scale dimension $d_{\UP}$ with
$v=10^{-3}$ is shown in Fig.~\ref{fig:sold}, where the numbers in
the plot stand for the values of $S_0$.
By the figure, we see that $S_0$ is increasing while $d_{\UP}$ is
decreasing. Furthermore, the value of $d_{\UP}$ for $S_0\sim {\cal
O}(10^3)$ can be somewhat larger, when $\lambda$ is away from unity.
It is clear that Sommerfeld factor could be as large as ${\cal
O}(10^{3})$ by the new force mediated by unparticle.

\begin{figure}[b]
\includegraphics*[width=3 in]{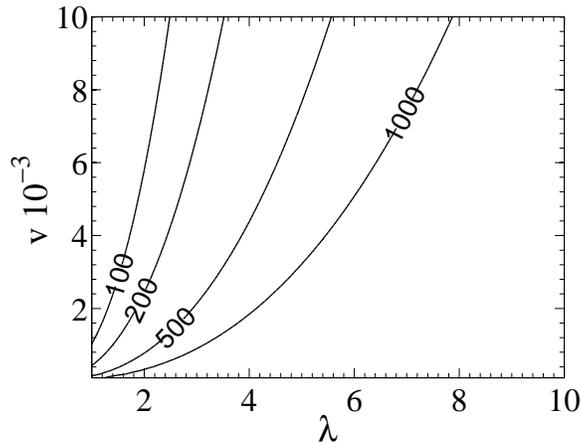}
\caption{Legend is similar to Fig.~\ref{fig:sold}, where
$d_{\UP}=1.1$ and the variables for the contour are $\lambda$ and
$v$ (in units of $10^{-3}$). }
 \label{fig:solv}
\end{figure}

Secondly, we also study the behavior of $S_0$ in $\lambda$ and
$v$ (in units of $10^{-3}$) by fixing the value of $d_{\UP}$. We
present the resultant contour with $d_{\UP}=1.1$ in
Fig.~\ref{fig:solv}. Clearly, with the value of $d_{\UP}$ that is
close to unity, $S_0$ can easily reach ${\cal O}(10^{3})$ even with
$\lambda\sim 1$ and $v\sim 10^{-3}$. On the other hand, $S_0$ of
${\cal O}(10^{3})$ can be preserved when $v$ and $\lambda$ both are
increasing simultaneously.

Finally, for understanding the dependence of speed of WIMP and
scale dimension of unparticle on Sommerfeld factor, we checked the
contour for $S_0$ as a function of $v$ (in units of $10^{-3}$) and
$d_{\UP}$ with $\lambda=1$.
And we found that Sommerfeld enhancement of ${\cal O}(10^2)$ for explaining
the excess of observed cosmic-ray through WIMP annihilation can
still be achieved with $v\sim {\cal O}(10^{-3})$ and $\lambda=1$ while
$1<d_{\UP}<1.1$.

As summary, inspired by the excess of electrons and/or positrons observed at
PAMELA, ATIC, Fermi-LAT and $etc.$, the issue of dark matter
annihilation for the solution is revived and studied broadly.
Besides a new mechanism is needed for the production of excessive
cosmic-ray, usually we also need a new force that interacts between
dark matter for overcoming the low cross section when dark matter
collides. For studying the possible new force, we have investigated
the impact of unparticle which is ruled by scale invariant.

We find that the power of the static unparticle potential in $r$,
resulted from the exchange of scalar unparticle, is non-integral
number and it depends on the scale dimension of unparticle,
expressed by $1/r^{2d_{\UP}-1}$. By the boundary condition for wave
function at $r=0$, the upper bound of scale dimension is found by
$d_{\UP}|_{\rm max} < 3/2$. By looking for the suitable boundary
condition and matching condition for the radial wave function at
$r\to 0$, we are led to the Sommerfeld factor for $s$-wave
collision, in which the result could return to that of Coulomb
potential. However, unlike Coulomb or Yukawa potential, the speed
dependence is related to scale dimension of unparticle and dictated
by $1/v^{3-2d_{\UP}}$.
Although the Sommerfeld factor is associated several parameters,
however, our results are only sensitive to the parameters $\lambda$,
$d_{\UP}$ and $v$. According to our numerical calculations, we
conclude that with the allowed range of free parameters, the
Sommerfeld enhancement induced by unparticle-mediated effects could
be ${\cal O}(10-10^{3})$ and the factor could provide the necessity for
enhancing the cross section of dark matter annihilation.
\\

\noindent[{\it Acknowledgments}]~~~~
This work for C.H.C. is supported by the National Science Council of R.O.C.
under Grant No: NSC-97-2112-M-006-001-MY3.
The work of C.S.K. is supported in part by Basic Science Research Program through the NRF of Korea
funded by MOEST (2009-0088395),  in part by KOSEF through the Joint Research Program (F01-2009-000-10031-0),
and in part by WPI Initiative, MEXT, Japan.

\end{document}